\documentclass{article}
\PassOptionsToPackage{numbers, compress}{natbib}

\PassOptionsToPackage{numbers}{natbib}
    \usepackage[final]{neurips_2020_ml4ps}

\usepackage[utf8]{inputenc} 
\usepackage[T1]{fontenc}    
\usepackage{hyperref}       
\usepackage{url}            
\usepackage{booktabs}       
\usepackage{amsfonts}       
\usepackage{nicefrac}       
\usepackage{microtype}      
\usepackage{graphicx}
\usepackage[para]{footmisc}
\usepackage{enumitem}

\title{Estimating Galactic Distances From Images Using Self-supervised Representation Learning}

\author{Md Abul Hayat \\
   University of Arkansas\\
   Fayetteville, AR 72701\\
   \texttt{mahayat@uark.edu}\\
   \And
   Peter Harrington \\
   Lawrence Berkeley National Laboratory\\
   Berkeley, CA 94720\\
   \texttt{pharrington@lbl.gov}\\
   \And
   George Stein \\
   Berkeley Center for Cosmological Physics \\
   Lawrence Berkeley National Laboratory\\
   Berkeley, CA 94720\\
   \texttt{georgestein@lbl.gov}\\
   \And
   Zarija Luki\'{c}$^*$ \\
   Lawrence Berkeley National Laboratory\\
   Berkeley, CA 94720\\
   \texttt{zarija@lbl.gov}\\
   \And
   Mustafa Mustafa\thanks{Corresponding authors} \\
   Lawrence Berkeley National Laboratory\\
   Berkeley, CA 94720\\
   \texttt{mmustafa@lbl.gov}\\
}

\begin{document}

\maketitle

\begin{abstract}

We use a contrastive self-supervised learning framework to estimate distances to galaxies from their photometric images. We incorporate data augmentations from computer vision as well as an application-specific augmentation accounting for galactic dust. We find that the resulting visual representations of galaxy images are semantically useful and allow for fast similarity searches, and can be successfully fine-tuned for the task of redshift estimation. We show that (1) pretraining on a large corpus of unlabeled data followed by fine-tuning on some labels can attain the accuracy of a fully-supervised model which requires 2-4x more labeled data, and (2) that by fine-tuning our self-supervised representations using all available data labels in the Main Galaxy Sample of the Sloan Digital Sky Survey (SDSS), we outperform the state-of-the-art supervised learning method.
\end{abstract}

\section{Introduction}

Sky surveys collect large amounts of data that can be used to glean information about the physical model of the universe and its evolution. One significant problem is determining 3D position of each galaxy, since its distance from the Earth is not directly measurable as are the two other spatial coordinates. Instead, one can measure the redshift of each galaxy ($z$) using its spectrum by looking for the departure of spectral lines from their rest-frame position (we will call this ``spec-z'').  This is precise, but is a very costly procedure, as taking a spectrum requires far longer exposure times than making a photo-image of a galaxy.  As a result, the vast majority of observed galaxies do not have spec-z; instead, their redshifts are estimated from photometric data \citep{Baum1962,salvato2019many}, i.e.~from images in a few passbands (think of them as the astronomical version of RGB color space, generally with $\sim$5 channels). Historically, these ``photo-z'' estimates commonly relied on template fitting \citep{Loh1986} or linear regression \citep{Connolly1995} as a function of the total flux of each galaxy in different passbands, and neglected structural details from the images themselves. Deep neural networks however, open up new and exciting possibilities for learning photo-z's from galaxy images using the limited number of spec-z labels.  Last year, a fully-supervised convolutional neural network (CNN) was used for this task \citep{pasquet2019photometric}, showing significantly lower errors than previous models. While encouraging, this approach requires lots of labeled data.

Recent research shows that self-supervised learning techniques enable CNNs to build meaningful visual representations without needing per-sample labels for any specific downstream task \citep{he2020momentum, chen2020simple, chen2020improved}. When labels are available for only a fraction of the full dataset, the representations built via self-supervised training can then be fine-tuned for the specific vision task, and this approach is able to significantly outperform fully-supervised training \citep{chen2020big}. Similarly, recent results in semi-supervised learning utilize a large body of unlabeled data to complement the labeled examples, and achieve state-of-the-art performance in image classification tasks as a result \citep{Xie_noisystuden2019, Touvron_traintest2020}.

In this work we demonstrate the advantage of self-supervised approaches for photometric redshift estimation using data from SDSS\footnote{https://www.sdss.org/}.
As a first step, we establish a new fully-supervised baseline on this data. Then,  using contrastive self-supervised pretraining, we show that such methods can build meaningful visual representations which are useful for a variety of tasks. By pretraining on a large corpus of unlabeled data followed by fine-tuning on labels, we can achieve the same performance as fully-supervised approaches which require 2-4x more labeled data. Finally, using the above approach and fine-tuning on the full labeled dataset, we establish a new image-based ML baseline on this dataset.

\section{Methods}

We use data from SDSS, a major sky survey conducted on a 2.5-meter telescope at Apache Point Observatory in New Mexico.
In total, it provides photometric observations of roughly 1 billion objects in the sky, and spectra for approximately 4 million objects (stars, galaxies, and quasars).

{\bf{Galaxy labels.}} We closely follow the process of \cite{pasquet2019photometric} in building the labeled portion of our training dataset to enable direct comparison to their results. Using their SQL query on the \url{skyserver.sdss.org/CasJobs} service, we pull samples from the Main Galaxy Sample in the 12th Data Release (DR12; \cite{alam2015eleventh}) of the SDSS, filtering for objects classified as \texttt{`GALAXY'} with dereddened petrosian magnitudes $\leq 17.8$ and spectroscopic redshifts $z \leq 0.4$. For us, this query returns 547,224 objects, of which 502,977 are unique to use as labeled training examples. Executing a similar query on the \texttt{`PhotoObjAll'} full photometric catalog of the SDSS, and removing duplicates which were already included in our spectroscopic sample, gives an additional body of unlabeled training examples. We imposed a cut at galactic latitude  $|b|=15^\circ$ to remove samples with high extinction near the galactic plane. We also excluded unlabeled samples with an estimated photometric redshift above 0.8 (as estimated by \cite{Becketal16photozbaseline}) to eliminate objects which are very likely too distant compared to the spectroscopic sample. After imposing these cuts, our dataset of additional photometric samples contains 1,194,779 objects and is used for pre-training. The labeled training and validation dataset have 399,984 and 102,993 images respectively.

\begin{figure}[t]
  \centering
    \includegraphics[width=0.49\textwidth, 
    trim={0.25cm 0.25cm 0.25cm 0.25cm}, clip]{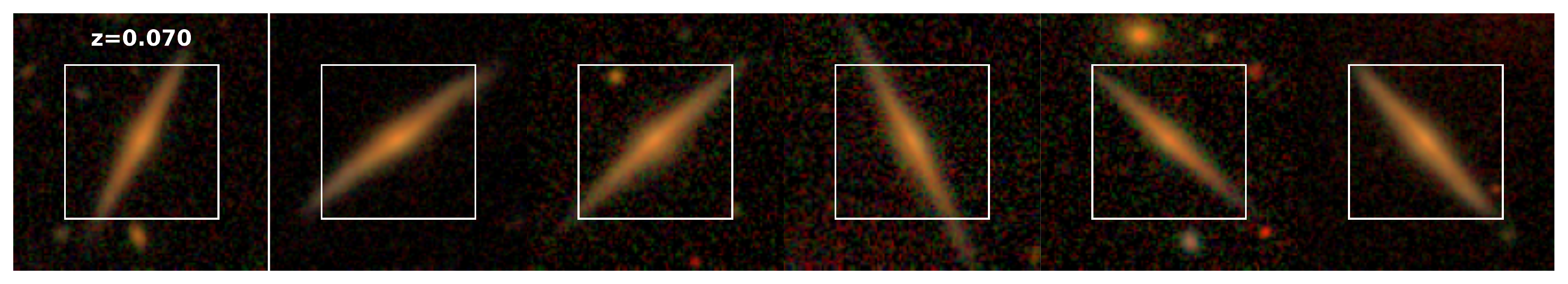}
    \includegraphics[width=0.49\textwidth, 
    trim={0.25cm 0.25cm 0.25cm 0.25cm}, clip]{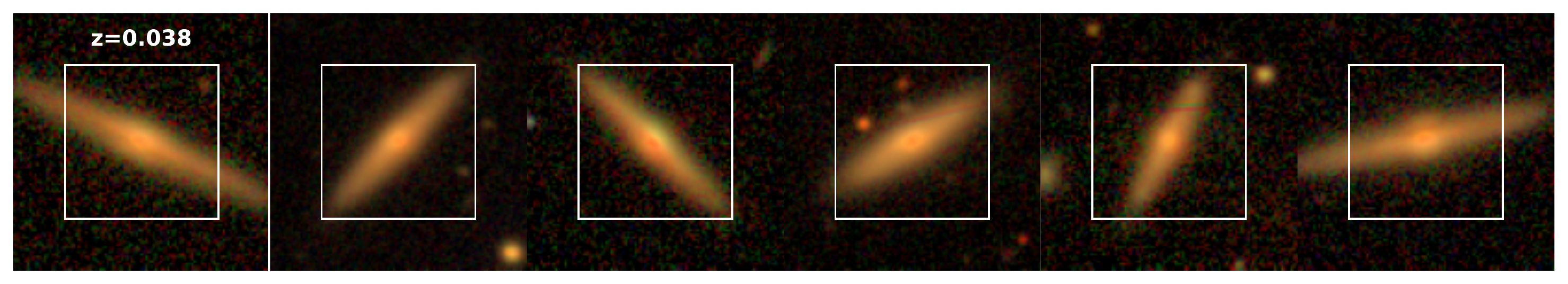}
    
    \includegraphics[width=0.49\textwidth, 
    trim={0.25cm 0.25cm 0.25cm 0.25cm}, clip]{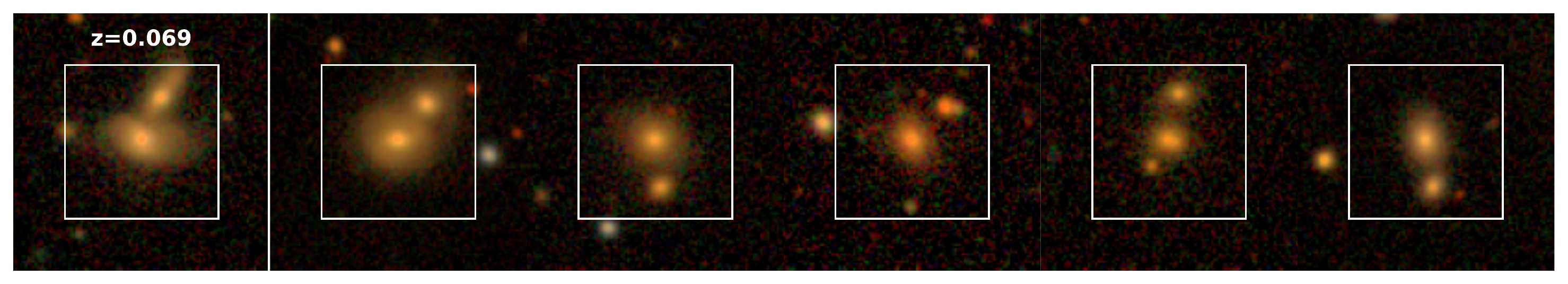}
    \includegraphics[width=0.49\textwidth, 
    trim={0.25cm 0.25cm 0.25cm 0.25cm}, clip]{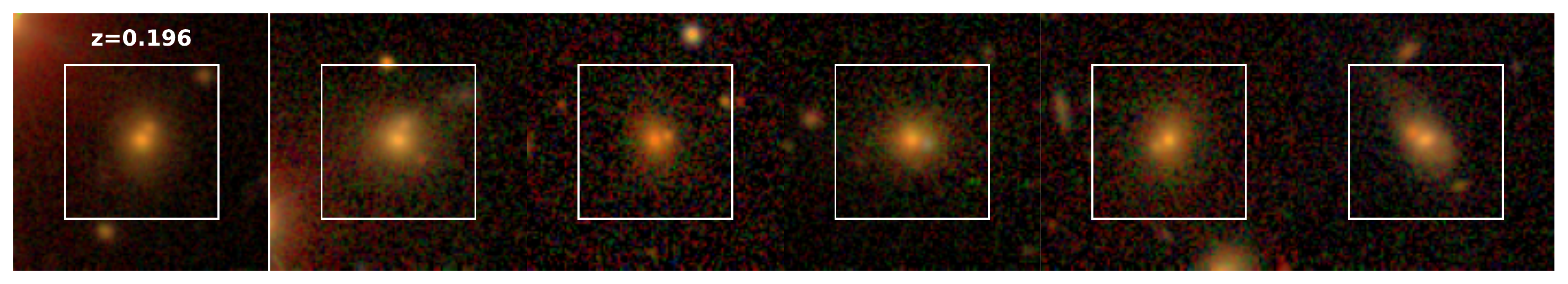}

    \includegraphics[width=0.49\textwidth, 
    trim={0.25cm 0.25cm 0.25cm 0.25cm}, clip]{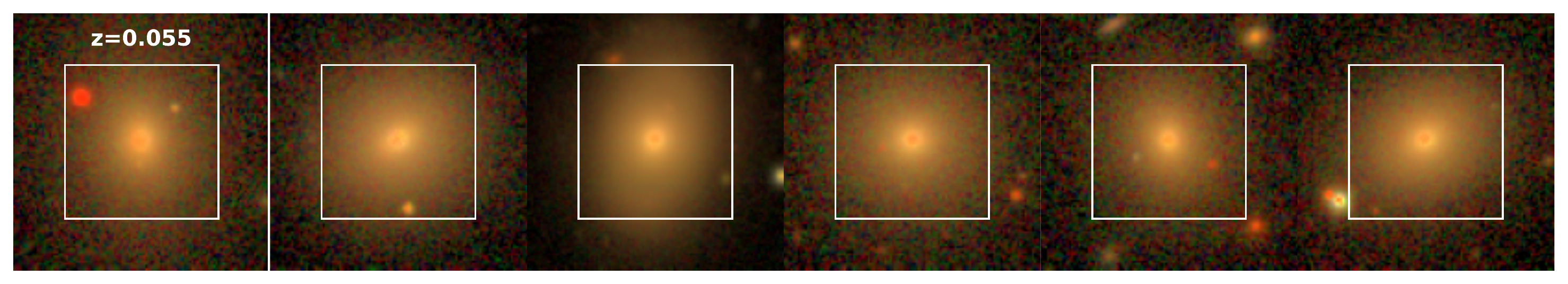}
    \includegraphics[width=0.49\textwidth, 
    trim={0.25cm 0.25cm 0.25cm 0.25cm}, clip]{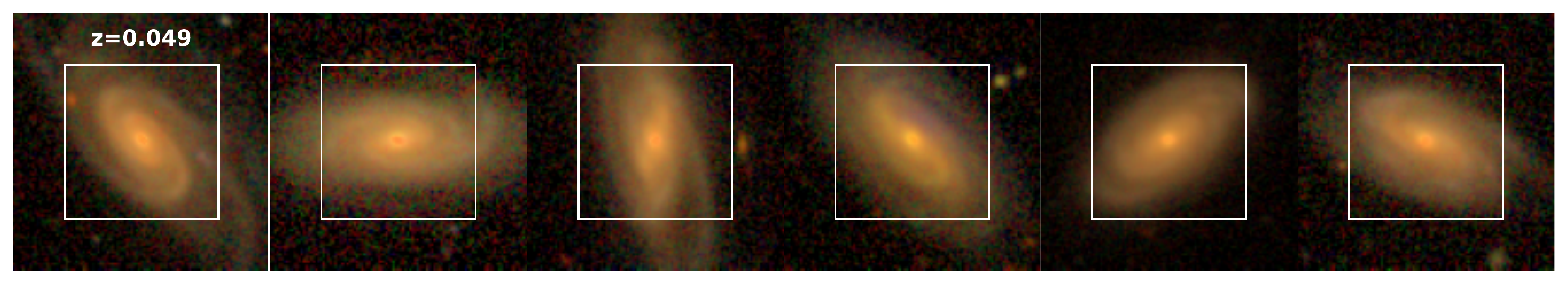}
    
    \includegraphics[width=0.49\textwidth, 
    trim={0.25cm 0.25cm 0.25cm 0.25cm}, clip]{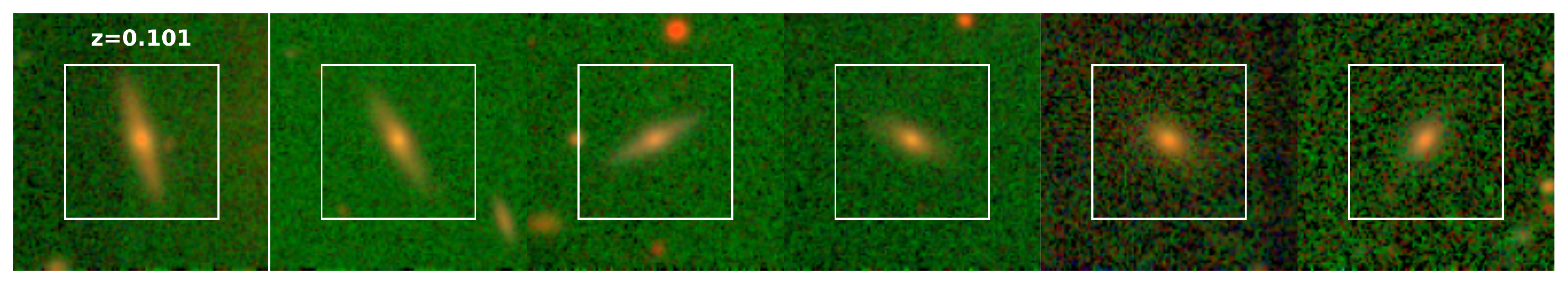}
    \includegraphics[width=0.49\textwidth, 
    trim={0.25cm 0.25cm 0.25cm 0.25cm}, clip]{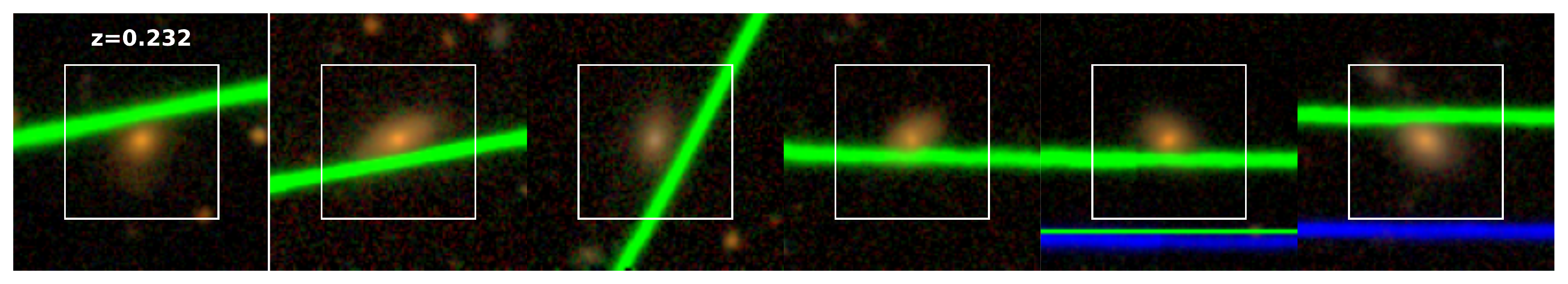}
    
    \includegraphics[width=0.49\textwidth, 
    trim={0.25cm 0.25cm 0.25cm 0.25cm}, clip]{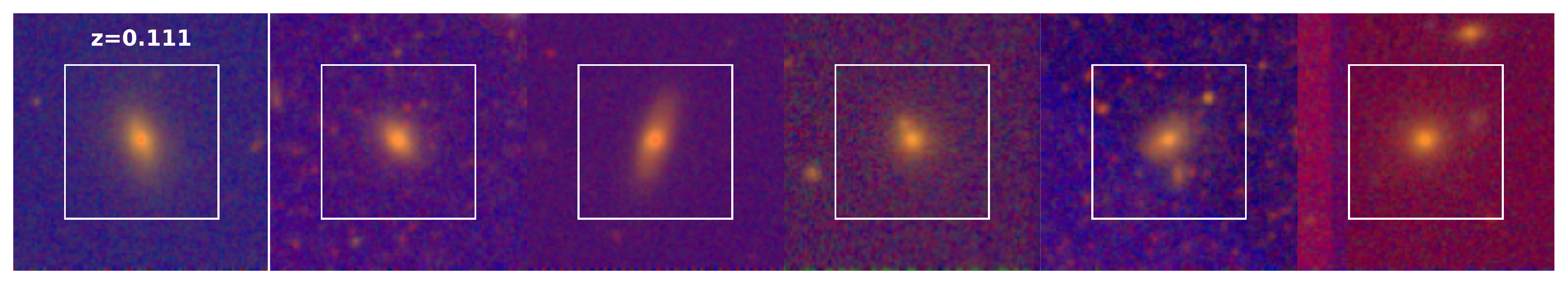}
    \includegraphics[width=0.49\textwidth, 
    trim={0.25cm 0.25cm 0.25cm 0.25cm}, clip]{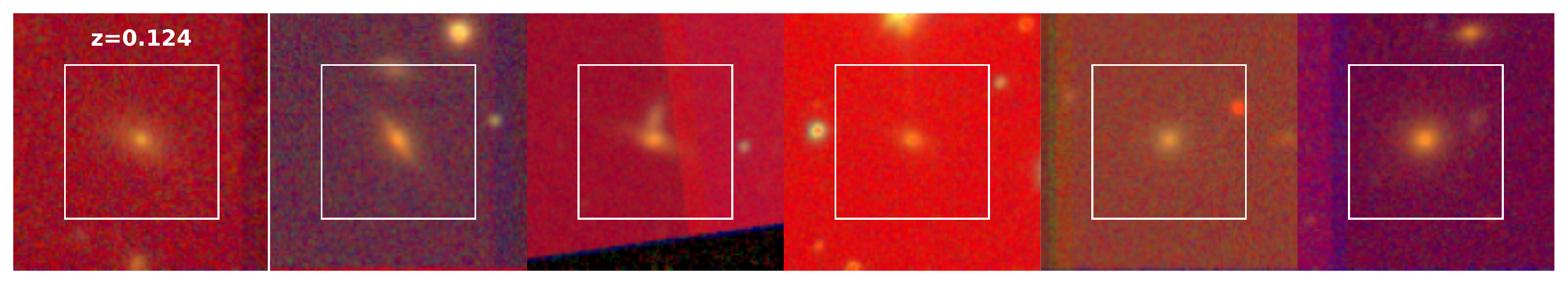}
    
\vspace{-7pt}
  \caption{Reference SDSS galaxies from the validation set (leftmost panels with redshift labels) and the most similar galaxies from the training set (following 5 panels) identified through a self-supervised similarity search. White squares outline the 64$^2$ pixels that are ``seen'' by the network.}
  \label{fig:sample_images}
\end{figure}

{\bf{Galaxy images.}} SDSS photometric images contain data in 5 passbands ($ugriz$), and come background-subtracted, but are not de-reddened to account for galactic extinction. We use the Montage\footnote{http://montage.ipac.caltech.edu/} tool to query the imagery catalog in SDSS Data Release 9 (DR9), based on equatorial coordinates for each object in our dataset. We sampled a $(0.012^\circ)^2$ patch of sky centered on each object, and projected onto a 2D image with $107^2$ pixels to ensure the pixel scale is as close as possible to the native pixel scale in the SDSS, 0.396 arcsec. In each image, we store the $u$, $g$, $r$, $i$, and $z$ passbands as 5 color channels. Note that during training of the self-supervised model we impose random rotations and random jitter to each image before cropping out the central portion as a data augmentation, so while our images contain 107 pixels per side, the CNNs in this work only view samples of size 64$^2$, which is consistent with \cite{pasquet2019photometric}. Galaxies with smaller redshift i.e. closer to our galaxy may have a larger shape that goes beyond size 64$^2$, but number of such galaxies being really small we limit the view to 64$^2$ for simplicity. Sample images (transformed into RGB color space) are shown in Fig.~\ref{fig:sample_images}.

{\bf{Self-supervised Learning.}}
Recent self-supervised models like SimCLR \cite{chen2020simple,chen2020big} and MoCo \cite{he2020momentum,chen2020improved} achieve success using a contrastive learning approach, which aims to build representations that are invariant under various transformations (e.g., random crops, Gaussian noise, or color jitter). These networks are trained with contrastive losses like NT-Xent as in \cite{chen2020simple} or InfoNCE in \cite{oord2018representation}, which try to maximize the similarity of two augmented versions of the same image while minimizing similarity with augmented versions of other images. 

In this work we use the MoCov2 framework \cite{chen2020improved}, which employs randomized data augmentations to produce two different views $x^q, x^k$ of each training image. 
In each training step, the differing views of each sample are passed through two independent networks, an encoder and a momentum encoder, which have identical architectures. This yields representations $q$ and $k^+$ which represent a ``positive'' pair of samples, while ``negative'' examples $k^-$ are retrieved from a queue containing previous outputs of the momentum encoder on other augmented images. The encoder network is trained via backpropagation using the InfoNCE loss
\begin{equation}
    L_{q,k^+,\{k^-\}} = -\log \frac{\exp(q\cdot k^+/\tau)}{\exp(q\cdot k^+/\tau) + \sum_{k^-} \exp(q\cdot k^-/\tau)},
\end{equation}
where, $\tau \in (0,1)$ is a temperature parameter. InfoNCE assigns higher similarity scores between the augmented
views of the same image compared to augmented views from other images. Then, the parameters $\theta_k$ of the momentum encoder network are updated using the encoder parameters $\theta_q$ with momentum parameter $m$ via
\begin{equation}
    \theta_k \leftarrow m\theta_{k}+(1-m)\theta_q. 
\end{equation}

The momentum update and use of a queue allow many negative examples to be seen throughout training without requiring massive batch sizes, making MoCo computationally efficient \cite{chen2020improved}. Following MoCov2, we set $m=0.999$ and use the ResNet50 architecture \cite{HeResNet} for the encoder and momentum encoder networks. However, we remove the first convolution and pooling layers in the ResNet and replace them with a single stride=1 convolution with 5 input feature channels to match the dimensionality of our dataset.

{\bf{Data Augmentations.}}
We create different views of training samples using the following data augmentations:
 \begin{itemize}[leftmargin=*]
     \itemsep0em 
    \item \textit{Galactic extinction:} To model the effects of foreground galactic dust, we introduce artificial reddening by sampling a $E(B-V)$ reddening value from $\mathcal{U}(0,0.5)$ and applying the corresponding per-channel extinction according to  the photometric calibration from \cite{schlafly2011measuring}.
    \item \textit{Random rotate:} The angle of rotation is sampled from $\mathcal{U}(0,2\pi)$. 
    \item \textit{Random jitter \& crop:}  Two integers are sampled from $\mathcal{U}(-7,7)$ to move (jitter) the center of the image along each respective axis, then the jittered image is center-cropped to size 64$^2$.
    \item \textit{Gaussian noise:} We sample a scalar from $\mathcal{U}(1,3)$ and multiply it with the aggregate median absolute deviation (MAD) of each channel (pre-computed over all training examples) to get a per-channel noise scale $\gamma_c$. Then, we introduce Gaussian noise sampled from $\mathcal{N}(0,\gamma_c)$ for each color channel.
\end{itemize}

\begin{figure}[t]
    \centering
    \includegraphics[width=0.95\textwidth]{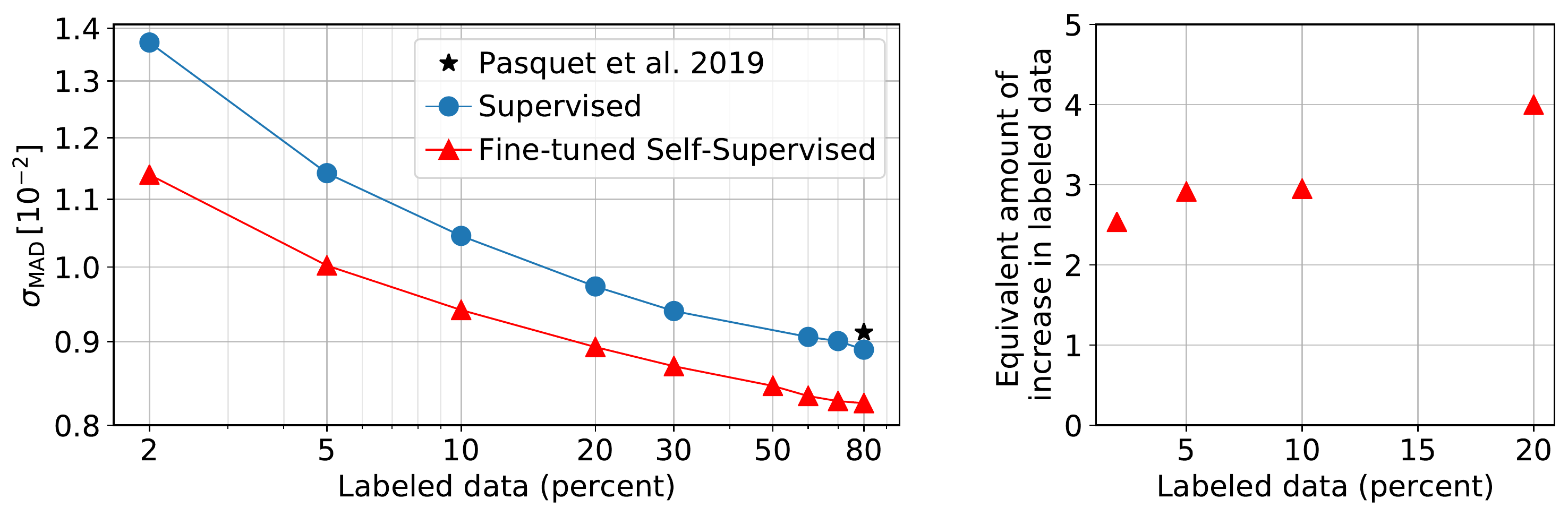}
    \caption{Our fine-tuned self-supervised approach outperforms a supervised ResNet50 at different fractions of training data (left). To achieve the same $\sigma_{\rm MAD}$, the amount of labeled data needed for the supervised approach is 2-4x higher than for the self-supervised approach (right).} 
    \label{fig:metrics}
\end{figure}

\section{Results}

To evaluate the quality of our redshift estimates, we follow the statistics used in the literature \cite{salvato2019many, pasquet2019photometric}.
The prediction residual is defined as $\Delta z = (z_{p}-z_{s})/(1+z_{s})$, where $z_p$ and $z_s$ correspond to photometric and spectroscopic redshifts respectively. From this, we compute the MAD deviation $\sigma_{\rm MAD}= 1.4826 \times \rm{MAD}(\Delta z)$ where MAD is the median of $|\Delta z - \mathrm{median}(\Delta z )|$, and the percent $\eta$ of ``catastrophic'' outliers with $|\Delta z| > 0.05$. To the best of our knowledge, \cite{pasquet2019photometric} provide state-of-the-art fully-supervised photometric redshift estimates, achieving $\sigma_{\rm MAD}=0.00912$ and $\eta = 0.31$ using all of their labeled data. Their network is trained to be a classifier over 180 bins of photometric redshifts linearly spanning $z=(0, 0.4)$. Their estimate is computed as $z_p = \mathbb{E}(z)$ using the probability density function given by the final softmax layer, so we use the same protocol.

To properly evaluate our ResNet-based self-supervised approach, we also train fully-supervised ResNet50s on de-reddened versions of our labeled samples. This is done with 2.5\%,  6.25\%, 12.5\%,  25\%,   37.5\%,  62.5\%,  75\%,   87.5\% and  100\% of the training data, with a fixed validation set. Our ResNet50, with 100\% of our labeled training data, sets a new benchmark on fully-supervised estimation of photometric redshift with $\sigma_{\rm MAD} = 0.00890$ (blue points in Fig.~\ref{fig:metrics}) and $\eta = 0.304$, a marginal improvement over the previous work.

{\bf{Self-supervised representations.}} After the contrastive learning phase, galaxies are passed through the encoder network to obtain their 128 dimensional contrastive loss vectors. By computing the dot products of a query vector with those of the unlabeled training samples, and sorting by the largest values, galaxies are returned sorted by their similarity to the query. These results are depicted in Fig. \ref{fig:sample_images}, where the top 3 rows show examples of common galactic morphologies, while the bottom 2 have identified a number of potential observational errors that unknowingly remained after the data quality cuts were imposed. Such observational errors could be caused by foreground sources like satellites or aircraft, or by cosmic rays, leading to artifacts like the bright green lines seen in some images. It is beyond the scope of this paper, but our methodology could be used for galaxy similarity identification (see, for example, reference \citep{Hocking2018}), or for anomaly detection.

{\bf{Self-supervised fine-tuning.}} After self-supervised pre-training on our unlabeled dataset, we partition the encoder network after the final \verb+avgpool+ layer (yielding a convolutional stage followed by a fully-connected layer). We fine-tune these sections on the de-reddened labeled data with a 10x smaller learning rate for the convolutional portion of the network, using the random rotations and random jitter-crop augmentations. We find that when fine-tuning our self-supervised representations on the fractions of the training dataset as explained above, we outperform the fully-supervised ResNet50 baseline for all fractions, as shown in Fig. \ref{fig:metrics}. Our self-supervised network, fine-tuned on 100\% of the labeled training data, outperforms both our fully-supervised baseline and the previous state-of-the-art result with $\sigma_{\rm MAD}=0.00825$ and $\eta=0.209$. Importantly, for the intermediate fractions of the training dataset we find that our self-supervised network achieves the equivalent accuracy of a fully-supervised network while using 2-4x less labeled data (right panel in Fig.~\ref{fig:metrics}). We have also confirmed that the prediction bias $\langle \Delta z \rangle$, defined as the mean of the residuals, is negligibly small in our work ($\langle \Delta z \rangle < 10^{-4}$) as it was in \cite{pasquet2019photometric}.

\section{Conclusion}
In this work we have presented a first effort in leveraging self-supervised learning to process unlabeled photometric data, showing that the resulting visual representations are semantically useful and can be fine-tuned to surpass the performance of fully-supervised photometric redshift estimation models. We demonstrate that our model performs well even in the case of limited data labels, which is extremely valuable given the relative cost of acquiring spectroscopic measurements for a large sample of galaxies. Upcoming large-scale sky surveys from the ground and space, like the Vera C.~Rubin Observatory\footnote{https://www.lsst.org/} and Euclid\footnote{https://www.euclid-ec.org/}, are projected to image tens of billions of galaxies over the next decade. With those surveys in mind, our results show great promise for self-supervised learning methods to assist in deriving more precise photometric redshift estimates, helping
address fundamental physics and cosmology questions on the nature and properties of dark energy, dark matter and gravity.

\section*{Broader Impact}
As a component of this publication we intend to (pursuant to SDSS approval) publicly release our processed version of galaxy images and redshift labels, as well as all networks used in this research.  Data we have pre-processed is immediately ready for different learning tasks and easy to manipulate with e.g.~python codes, which is not the case with the more complex SDSS data server.  Our full code release makes it straightforward to reproduce our results and build upon it.  This reduces the entry threshold for future machine learning explorations with this interesting data set, especially for researchers with no background in astronomy, and for students who are not in top-tier research universities.

\bibliographystyle{abbrvnat}
\bibliography{ref}

\end{document}